\documentclass{phbauth}        
\usepackage{graphicx}

\begin{document}

\begin{frontmatter}

\title{Scanning Tunneling Spectroscopy in MgB$_2$}

\author[address1]{P. Martinez-Samper, J.G. Rodrigo, G. Rubio-Bollinger, H. Suderow, S. Vieira}
\author[address2]{S. Lee, S. Tajima}

\address[address1]{Laboratorio de Bajas Temperaturas, Depto. Fisica de la Materia Condensada, Instituto de Ciencia de Materiales N. Cabrera, Universidad Autonoma de Madrid, E-28049 Madrid, Spain*}

\address[address2]{Superconductivity Research Laboratory, ISTEC, 1-10-13, Shinonome, Koto-ku, Tokyo, 135-0062 Japan}

\thanks[thank1]{
* Grupo Intercentros de Bajas Temperaturas, ICMM-CSIC and LBT-UAM.\\
Corresponding author: sebastian.vieira@uam.es}

\bigskip

\begin{abstract}
We present extensive Scanning Tunneling Spectroscopy (STM/S)
measurements at low temperatures in the multiband superconductor
MgB$_2$. We find a similar behavior in single crystalline samples
and in single grains, which clearly shows the partial
superconducting density of states of both the $\pi$ and $\sigma$
bands of this material. The superconducting gaps corresponding to
both bands are not single valued. Instead, we find a distribution
of superconducting gaps centered around 1.9mV and 7.5mV,
corresponding respectively to each set of bands. Interband
scattering effects, leading to a single gap structure at 4mV and a
smaller critical temperature can be observed in some locations on
the surface. S-S junctions formed by pieces of MgB$_2$ attached to
the tip clearly show the subharmonic gap structure associated with
this type of junctions. We discuss future developments and
possible new effects associated with the multiband nature of
superconductivity in this compound.

\end{abstract}

\begin{keyword}
MgB$_2$, superconductivity, tunneling spectroscopy, local
probes\end{keyword}

\end{frontmatter}

\section{Introduction}

The surprising discovery by Nagamatsu et al.\cite{Nagamatsu01} of
superconductivity at 40 K in the previously well known material
MgB$_2$, which is particularly simple from the structural point of
view, attracted immediately the interest of many research groups
in the world. Soon after its discovery, several experiments
appeared in the electronic archive cond-mat, and subsequently in
conventional journals, revealing some microscopic aspects of the
superconductivity in this compound that turned out later on to be
essential for its understanding. The boron isotope effect,
discovered by Bud'ko et al.\cite{Budko01} pointed towards a phonon
mediated pairing interaction, and the tunneling spectroscopy,
measured with a Scanning Tunneling Microscope (STM/S), showed a
clean BCS density of states with an intriguing low value for the
ratio $2\Delta/k_BT_c=1.2$ \cite{Rubio01}, as compared to the
value expected from simple BCS theory (3.53). First principles
investigations of the electron-phonon coupling concluded that two
gaps, related to the two different sets of bands of the Fermi
surface would exist in this superconductor\cite{Liu01}, and that
the gap reported in Ref.\cite{Rubio01} could be associated to one
of these sets of bands. Within another interesting approach,
Bascones and Guinea \cite{Bascones02} considered the effect of the
surface as a topological defect in two band superconductors to
explain the observation of a single gap structure in some early
tunneling
experiments\cite{Rubio01,Karapetrov01,Schmidt01,Sharoni01}.

Two band superconductivity was theoretically proposed by Suhl,
Matthias and Walker in 1959\cite{Suhl59}, and has been discussed
since then in relation to other materials. For instance, important
tunneling experiments were made by Binnig et al. in 1980
\cite{Binnig80} on Nb doped SrTiO$_3$. These authors were able to
measure in a very neat way the two quasiparticle peaks in the
superconducting tunneling density of states due to the opening of
two gaps corresponding to the two bands of the Fermi surface of
this compound, filled one after the other by the Nb doping. The
results were compared to reduced SrTiO$_3$, where only a broadened
single peak structure is observed, together with a four times
larger residual resistivity. Interband scattering due to an
increasing amount of defects or impurities is indeed expected to
gradually merge multigap features in the density of states into
the typical BCS single gap curve\cite{Sung67,Golubov97,Mazin02b}.

Even if first measurements of the tunneling density of states
reported the observation of a single gap structure corresponding
to different values of the superconducting gap (with
$2\Delta/k_BT_c$ between 3 and 4
\cite{Karapetrov01,Schmidt01,Sharoni01}), more recent experiments
\cite{Schmidt02,Iavarone02,Giubileo01,Badr02,Eskildsen02} are in
good agreement with the model proposed in \cite{Liu01}. Therefore,
we can now state that tunneling spectroscopy experiments made in
this material support well the two gap scenario. As it is well
known, other very powerful macroscopic thermal and magnetic
measurements, as well as microscopic spectroscopic probes, which
will be treated in detail by other authors in this volume,
produced more and more compelling evidence in this direction (see
e.g.
\cite{Szabo01,Naidyuk01,Bouquet01,Yang01,Wang01,Manzano02,Quilty02,Tsuda01}).

Previously existing band structure calculations \cite{Armstrong79}
were reconsidered in order to understand the microscopic nature of
the superconducting state\cite{Kortus01,An01,Brinkman02}. Precise
calculations based on the strong coupling formalism of Eliashberg
have been recently made by the authors of Ref.\cite{Choi02}. It
has been shown, and confirmed by comparing de Haas van Alphen
experiments with the band structure calculations
\cite{Yelland01,Mazin02}, that the electron phonon interaction
varies strongly on the Fermi surface. Mainly four sheets are
found, two of them being near cylinders due to the two dimensional
antibonding states of boron p$_{xy}$ orbitals, and two three
dimensional sheets from the $\pi$ bonding and antibonding states
of the boron p$_z$ orbitals\cite{Yelland01,Uchiyama01}. The
superconducting gap $\Delta(\overrightarrow{k})$ has been
calculated as a function of temperature and magnetic field and is
non zero everywhere on the Fermi surface. However, its size
changes strongly clustering around two main values related to the
$\sigma$ and $\pi$ sheets\cite{Choi02}. The smaller value, which
was measured in our first experiments \cite{Rubio01}, appears to
be related to the three dimensional $\pi$ sheets, and the larger
one to the two dimensional $\sigma$ sheets.

We can therefore affirm that today it is widely accepted that
MgB$_2$ is a multiband superconductor where the basic physics can
be theoretically understood in a very sound way. This opens a new
and interesting challenge to improve the accuracy of the
experimental results and push the theoretical developments towards
more and more narrow limits. The search for new phenomena
associated with two band superconductivity appears especially
promising. With respect to other known multiband superconductors,
as e.g. doped SrTiO$_3$\cite{Binnig80} or the nickel
borocarbides\cite{Canfieldgeneral}, MgB$_2$ has the considerable
advantage of having a high critical temperature and a relatively
simple Fermi surface.

Local vacuum tunneling spectroscopy at low temperatures, made
possible with the use of the STM/S, is a very promising technique,
as it gives a direct measurement of the local density of
states\cite{Chen}. This is even more useful in the case of
materials, as MgB$_2$, where the superconducting gap is strongly
$\overrightarrow{k}$ dependent. The use of an atomically sharp
counter-electrode whose (x,y,z) position can be changed at will
in-situ, gives in principle access to detailed information about
the superconducting density of states in very different regions of
the Fermi surface. However, to get fully reliable data we need to
reduce as much as possible all pair breaking effects acting on the
surface of the condensate. Careful filtering of all wires arriving
to the sample turns out to be extremely important. The measured
spectra most accurately approach the local superconducting density
of states (LDOS) of the sample when the measurement is made near
T=0K to reduce thermal smearing. Obviously, characterizing and
controlling the surface of the sample is of extreme importance for
a proper use of this technique.

We should remark that, although other type of spectroscopic
measurements provide very useful information about the density of
states, the STM/S has the advantage of being a direct probe where
a very high resolution in energy can be achieved by measuring at
sufficiently low temperatures. For example, photoemission
spectroscopy also gives a direct measurement of the density of
states in different parts of the Fermi surface, but present
experiments \cite{Tsuda01} have much less resolution in energy
than STM/S. Point contact spectroscopy has been widely used in the
case of MgB$_2$ (see e.g. \cite{Szabo01,Naidyuk01}), and it is
also a very powerful, but less direct probe of the density of
states.

Recently, small single crystals of MgB$_2$ have been grown in a
few laboratories\cite{Lee02,Karpinski02,Machida02}, and some
results on tunneling spectroscopy have been
reported\cite{Eskildsen02}. At zero magnetic field, STS in the c
direction at low temperatures has been found in agreement with our
early result on single grains of this material\cite{Rubio01}.
However, to get a proper fit to BCS theory a pair breaking
parameter and a finite gap distribution, which both account for
experimental smearing, need to be introduced\cite{Eskildsen02}.
The observation of a hexagonal vortex lattice is a new result,
although more experiments are needed to understand the microscopic
aspects of the Shubnikov phase in this multiband
superconductor\cite{Eskildsen02}. Other STM/S experiments at 4.2 K
in oriented thin films, detailed in another contribution to this
volume, have also found two gap features associated to different
tunneling directions\cite{Karapetrov01}.

Several questions remain open which need to be addressed using low
temperature STM/S. For example, given that this material is a
multiband superconductor, how do the multiple gaps reflect in the
tunneling characteristics at a given point on the surface? How can
we extract information about the dominant gap structure in this
material? How shall we deal with and identify the effect of
interband scattering? For instance, would it be possible to find a
region of the surface where a single gap is observed? A zero bias
peak due to Andreev bound states, which could be related to
complex phase relations between the Cooper pair wavefunction in
the different bands, has been predicted in multiband, s-wave,
superconductors\cite{Voelker02}. Are there hints of such a peak in
STS experiments? On the other hand, S-S junctions (which can
sometimes be found in STM/S experiments) in conventional single
gap superconductors show a series of small features in the
conductance below the superconducting gap due to multiple Andreev
reflection processes (subharmonic gap structure). At which voltage
should we expect these features to appear in two gap
superconductors? Actually, how do multiple Andreev reflections
occur between two multiband superconductors?.

In this paper we present some of our STM/S research in MgB$_2$,
made after the first results published in \cite{Rubio01}, and try
to give insight into some of the above mentioned opened questions.
New data show in both, single crystals and small grains, the same
reproducible behavior independent of the sample preparation
method. The two gap structures are identified in most typically
found spectra. We also discuss some serendipitous events where a
piece of MgB$_2$ accidentally attached to the tip permits the
observation of clear S-S tunneling features, including the first
(to the best of our knowledge) observation of the subharmonic gap
structure.

\section{Experimental and sample preparation}

All cryostats used to measure the data presented here (a $^4$He, a
$^3$He and a dilution refrigerator) are equipped with similar,
home built, STM set-ups and electronics which have been carefully
filtered to produce a clean electromagnetic environment in the STM
chamber. Measurements in Pb, Al and NbSe$_2$ were used to test the
energy resolution of all set-ups\cite{Suderow02b,Suderow02}. All
STM's feature a very high mechanical stability and macroscopic
in-situ positioning capabilities (x-y table) in a 2x2 mm$^2$
window. We have measured single crystals and small grains prepared
in different ways. The single crystals were grown under pressure
as described in \cite{Lee02} and are needle like with typical
sizes of 0.1x0.06x0.7 mm$^3$ and the c axis along its long
direction. Samples were introduced (following its long axis) in a
small hole made on a sample holder (a flat gold surface) and glued
with a small drop of conductive silver epoxy, which was previously
put into the small hole. About half of the samples remained
outside the hole. Once a sample was mechanically fixed to the
sample holder, a small clean blade was used to cut it, and the tip
was positioned (with the help of an optical microscope and the
in-situ positioning capabilities of our STM) on top of the
remaining freshly broken surface. The whole set-up was then cooled
as fast as possible. The tunneling conditions were always good,
leading to reproducible images and spectra over the whole surface.
However, the topography is complex, showing a large amount of
inclined small terraces at the nanoscopic scale, observed with STM
and AFM, indicating that we could not find any cleaving plane. SEM
images at a larger scale also showed a terraced surface. This
means that the orientation of the surface at the length scale of
the STM experiment is unknown and does not coincide with the
breaking plane. The small grains were prepared following the
procedure described in Ref.\cite{Rubio01}. After dispersing them
in an acetone bath in ultrasound, a small drop was deposited onto
a flat gold surface, the liquid was evaporated at 80$^\circ$C and
the grains were pressed into the gold surface using a small
synthetic ruby. A subsequent bath in acetone and ultrasound
removed the loosely fixed grains. Following this process, some of
the samples were heated in a Ar atmosphere of 600mbar at
200$^\circ$C during two hours to promote gold diffusion and
improve the mechanical fixing of the grains.

We made an extensive topographical characterization of the surface
at all locations where we took conductance versus bias voltage
curves. However, here we report only about the most characteristic
measured spectra (STS), because we cannot clearly associate the
features of the spectra presented here to specific features of the
surface topography.

\section{Results and discussion}

\subsection{$\pi$ and $\sigma$ bands}

\begin{figure}[btp]
 \begin{center}\leavevmode
 \includegraphics[width=0.8\linewidth]{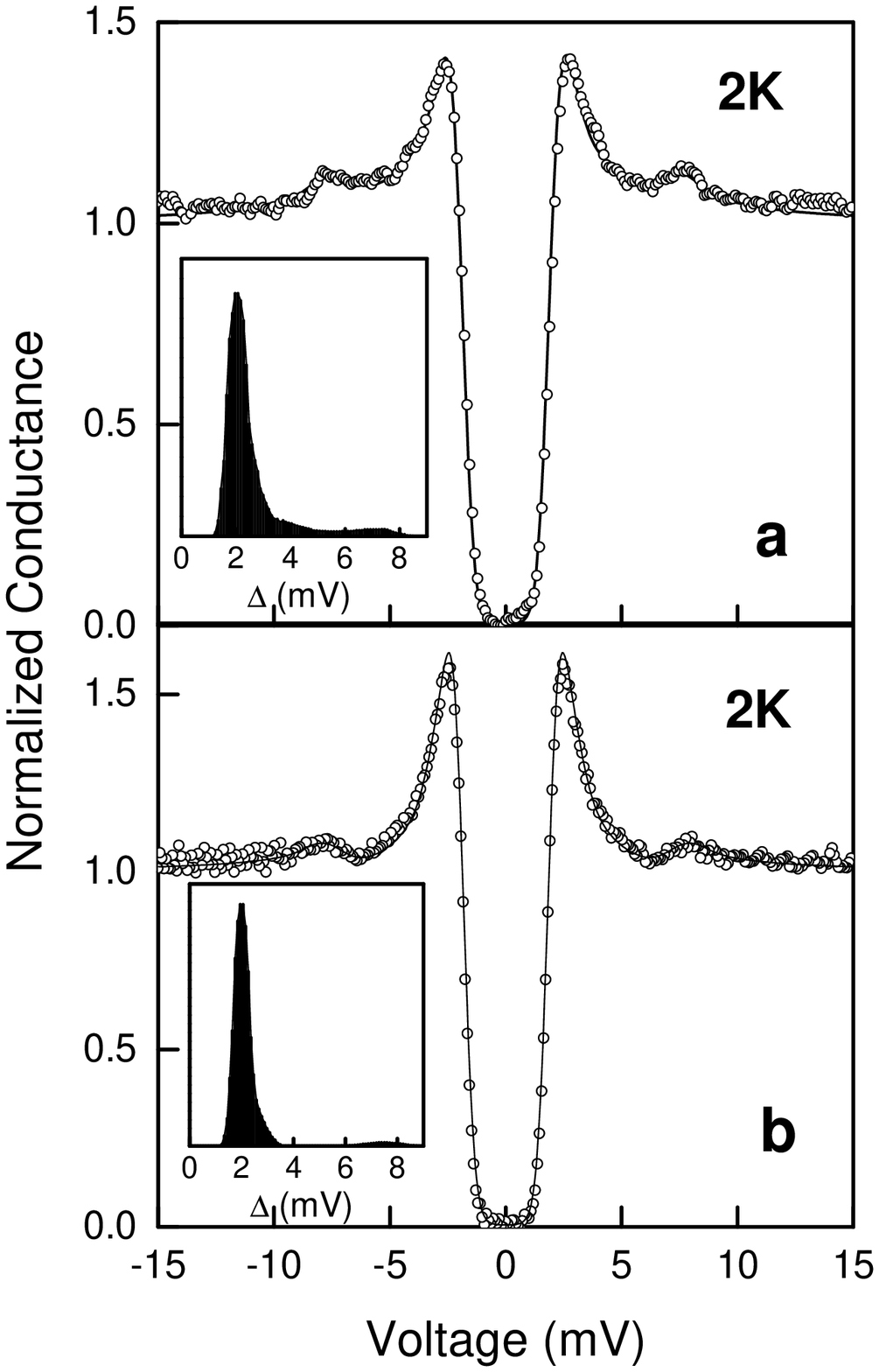}
\caption{Tunneling spectroscopy of MgB$_2$ at 2K (tunneling
resistance $R_N=1M\Omega$), measured in a single crystal (a) and
in a grain (b). The insets represent the (normalized) distribution
of values of the superconducting gap used to fit the experiment
according to the procedure described in the text (solid line:
model, circles: experiment). At this precise location, the
electrons contributing to the tunneling current come mostly from
the $\pi$ band. For clarity, a reduced number of the whole set of
experimental points is plotted in this and the following
figures.}\label{Fig1}\end{center}
\end{figure}

In this paper we report about many experiments made in
macroscopically different places of a large number of samples,
prepared in different conditions. The goal is to show the behavior
which, to our understanding, best reflects the intrinsic
properties of this superconductor. Millions of tunneling
conductance versus bias curves were measured. Most of them were
taken in points with good tunneling conditions (reproducible
imaging as a function of the bias voltage and work function of
several tenths of eV to several eV), and the results (measured in
all samples and preparation conditions) are comprised within two
limiting behaviors. Before discussing them we should remark that
an STM/S experiment always probes a part of the Fermi surface,
depending on the particular atomic configuration of tip and sample
surface\cite{Chen,Silkin01}. Therefore, a multiband material as
MgB$_2$ is expected to show a wide variety of behaviors, because,
depending on the location, one will probe different parts of the
Fermi surface.

In Fig.1 we show the tunneling conductance obtained in a small
grain and in a single crystal respectively. These curves show
clearly the partial density of states corresponding to the three
dimensional $\pi$ band. However, small bumps around 7.5mV can also
be observed. Fig.2 shows the other limiting behavior found in our
experiments, where the small bumps appearing in Fig.1 have evolved
into well developed peaks. Although these spectra are clearly
measured in both single crystals and grains, the curves of Fig.1
are the ones which appear most frequently on the surface of the
measured MgB$_2$ samples.

\begin{figure}[btp]
 \begin{center}\leavevmode
 \includegraphics[width=0.8\linewidth]{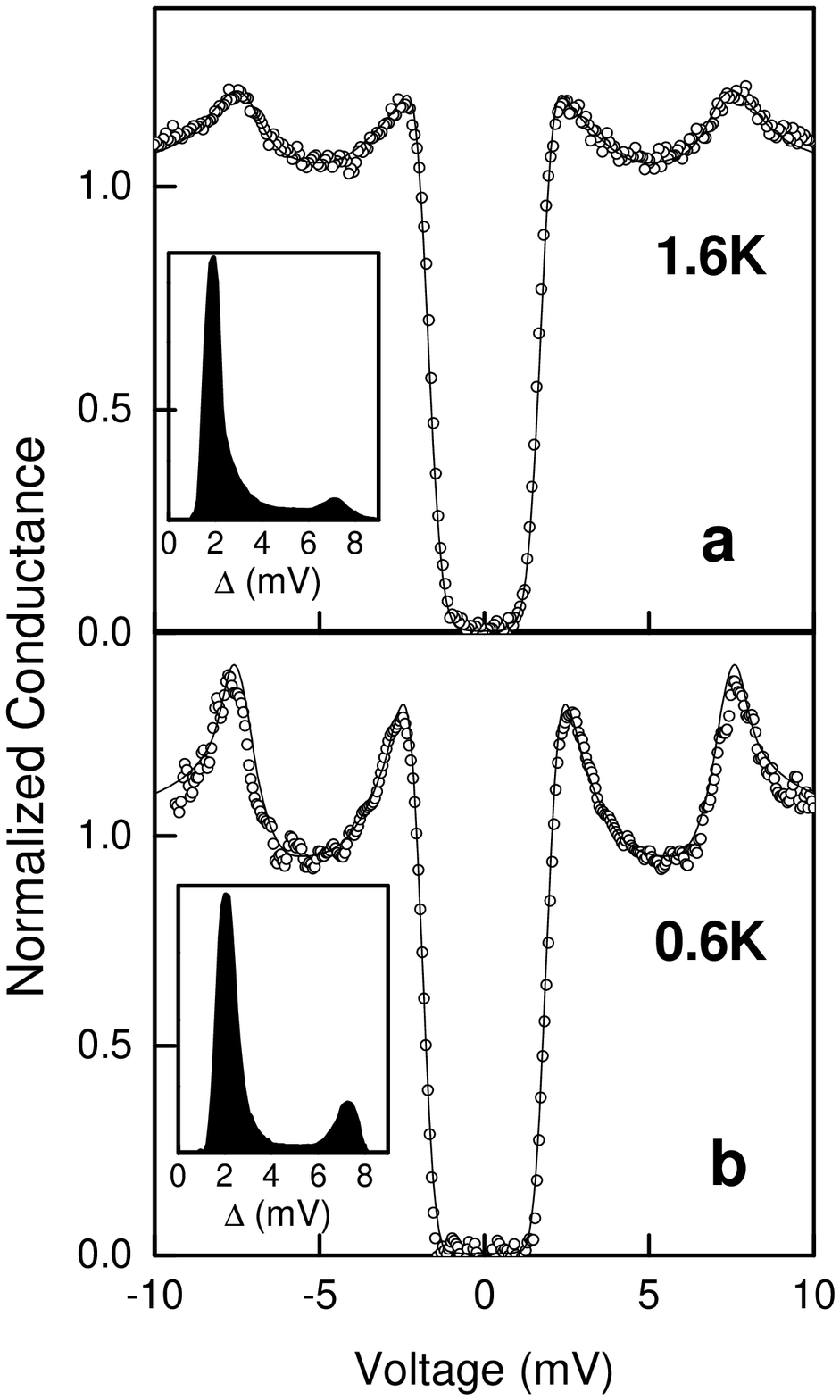}
\caption{Tunneling spectroscopy curves (circles) of MgB$_2$
obtained in a single crystal (a) and in a grain (b)
($R_N=1M\Omega$), together with the distribution of values of the
superconducting gap used to fit the experiment (lines). At this
precise location, the electrons contributing to the tunneling
current come both from the $\pi$ and $\sigma$ bands.
}\label{Fig2}\end{center}
\end{figure}

From these observations we can discuss the properties of the
superconducting gap in both bands. First, we should note that the
density of states within the smallest superconducting gap is
always zero. This implies that the pair breaking term
$\Gamma$\cite{Dynes78} is of no use to obtain an adecuate fit to
the experiments.

To obtain an idea of the contribution to the tunneling current
from the different bands, we phenomenologically model the measured
superconducting density of states by the sum over a normalized
distribution of BCS densities of states corresponding to a
distribution of values of the superconducting gap $\Delta$. As
shown in Fig.1, even when we deal with the simplest and most
frequent situation, i.e. a single gap structure around 1.9mV
corresponding to the $\pi$ band, a single valued gap is not enough
to obtain a the best fit to the experiment. The observed
quasiparticle peaks are not as high as expected from most simple
s-wave BCS theory. They are smaller because of a finite anisotropy
that gives a finite distribution of values of the superconducting
gap and therefore wider quasiparticle peaks. A gap distribution
due to the gap anisotropy smears out the coherence peaks but the
tunneling characteristics are flat-bottomed as observed in the
experiments. The situation is similar to the case of the
anisotropic superconductors NbSe$_2$ and the nickel borocarbide
materials \cite{Hess90,Suderow01a,Martinez02}, where the tunneling
characteristics taken at low temperature also show a zero density
of states at low energies but broadened quasiparticle peaks.

It is immediately apparent from our results that the tunneling
current comes mostly from the $\pi$ band electrons, because the
contribution to the partial density of states of the $\sigma$ band
gaps needed to explain our results is always relatively small.
When its intensity increases the contribution to the tunneling
current from electrons with $\overrightarrow{k}$ vectors
perpendicular to the c axis is also increasing.

Although our curves clearly indicate that the superconducting gap
is not single valued in both the $\pi$ and $\sigma$ bands, we must
emphasize that the phenomenological model used to fit the curves
here gives a very qualitative idea of the different contributions
to the tunneling current found in each location. Therefore, the
curves shown in the insets of Figs.1,2,3 and 5 are given to
illustrate the different behaviors found in different locations,
but they definitely do not give the actual distribution of values
of the superconducting gap corresponding to each band measured in
these locations. For instance, we believe that the filling of
region between the $\pi$ and $\sigma$ band gaps in our approach is
due to interband scattering effects.

\subsection{Interband scattering effects}

\begin{figure}[btp]
 \begin{center}\leavevmode
 \includegraphics[width=0.8\linewidth]{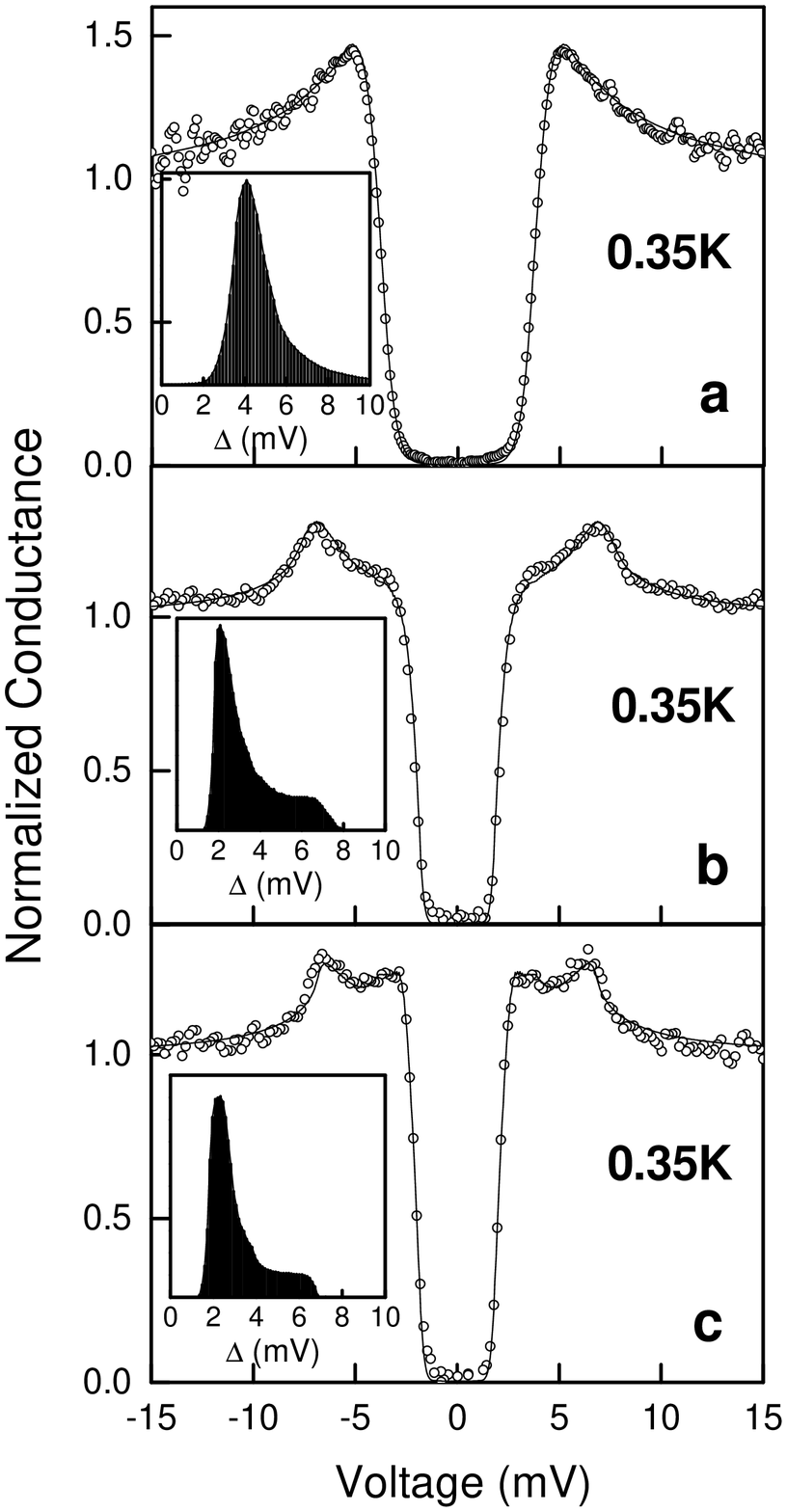}
\caption{Different spectra taken at different positions on a
single grain (circles), and its associated fits (lines and inset).
In (a) we observe a single, broadened quasiparticle peak, whereas
(b) and (c) show intermediate situations between those presented
in figs.1 and 2. We associate this behavior, which is also
observed in single crystalline samples, to interband scattering
effects at the surface of MgB$_2$ (see text).
}\label{Fig3}\end{center}
\end{figure}

As a matter of fact, it is difficult to understand in simple terms
the very robust superconducting properties within the two gap
model \cite{Mazin02b}. Indeed, multiband superconductivity is
expected to be sensitive to non magnetic defects or impurities, as
they allow for interband transitions, increasing the smaller gap,
decreasing the larger gap and leading therefore to a smaller
critical temperature\cite{Sung67,Golubov97,Mazin02b,Brinkman02}.
However, even bad quality samples of MgB$_2$ do not show big
changes in the superconducting critical temperature, and also show
both superconducting gaps with values which do not differ
significantly from the best samples. The theoretical work in
Ref.\cite{Mazin02b} deals with this apparent contradiction by
proposing that interband scattering is particularly small in this
compound. It would be highly desirable to obtain experimental
evidence that changes in the superconducting density of states due
to defects or impurities can indeed be observed as it is an
essential ingredient of multigap
superconductivity\cite{Sung67,Golubov97,Mazin02b,Brinkman02}.
Recent bulk specific heat measurements \cite{Wang02} on irradiated
samples have not been conclusive. The calculations of
Ref.\cite{Bascones02} indicate however that the surface can
produce in some cases interband scattering effects that do not
influence the bulk properties and which can be studied with STM/S.

We can find areas on the surface of MgB$_2$ in which the
conductance appears as shown in Fig.3a and which can be fitted
with distributions of values of $\Delta$ centered around 4mV, i.e.
between $\Delta_{\pi}$ and $\Delta_{\sigma}$. The superconducting
features of these spectra disappear (the spectra become completely
flat) around 20K. Note that the shape of this curve is in complete
disagreement with possible S-S tunneling features, discussed
further on and measured in Ref.\cite{Schmidt01}. We believe that
this behavior (also observed in some single crystalline samples)
is due to enhanced interband scattering in some regions on the
surface.
 The value of the superconducting gap and the
associated decrease in the local critical temperature agrees well
with the predictions of Refs.\cite{Mazin02b,Brinkman02}. Note that
this is a surface effect, related to the intrinsic multigap nature
of superconductivity in this compound, and which is clearly not
found in bulk samples\cite{Bascones02,Wang02}. Similar
observations have been done with STM/S in the non-magnetic nickel
borocarbides, where the surface shows a very irregular topography,
as in MgB$_2$\cite{Martinez02}. Note that point contact
experiments made in MgB$_2$ \cite{Naidyuk01} also show clearly a
single gap feature with $\Delta$ around 4mV in some measured
curves, in agreement with our results and discussion. Possibly, as
noted in Ref.\cite{Iavarone02}, early tunnel experiments showing a
smeared single gap structures around 3-4mV could be interpreted in
a similar way as the spectra shown in Fig.3a.

In Figs.3b and c we show other characteristic behaviors. Within
our simple phenomenological fits, the interband scattering fills
the intermediate region between both bands, as shown in the
insets. However, the features due to the $\pi$ and $\sigma$ bands
are still observed.

\subsection{S-S junctions}

\begin{figure}[btp]
 \begin{center}\leavevmode
 \includegraphics[width=0.8\linewidth]{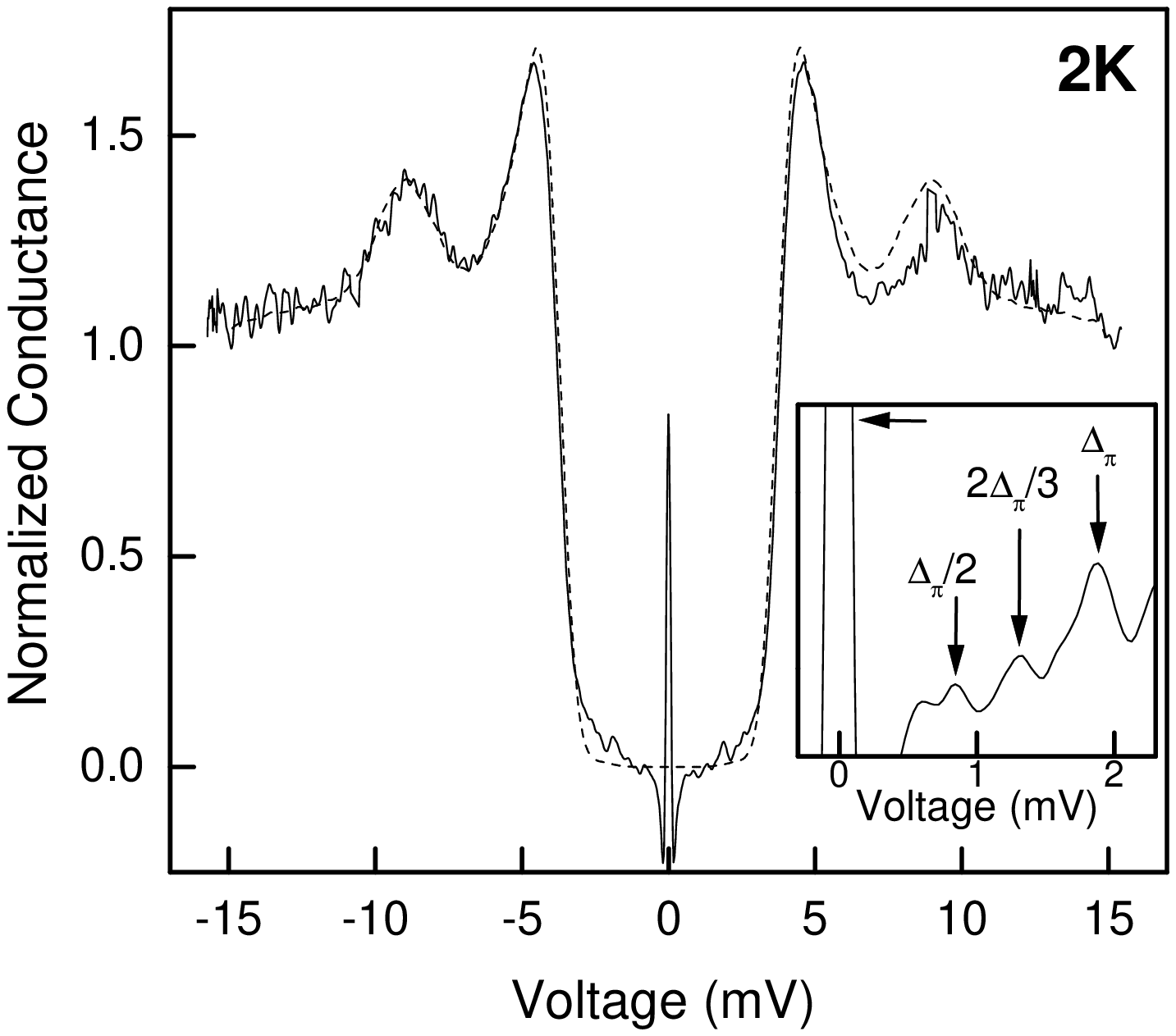}
\caption{Subharmonic gap structure observed in spectra
characteristic for tunneling between two single grains of MgB$_2$,
taken at 2K. The inset shows a zoom over the relevant voltage
region with arrows at each peak corresponding to a submultiple of
2$\Delta_{\pi}$, and at the peak originating in the conductance
due the Josephson effect (arrow to the left). Note that the
measured negative resistance around zero bias is  a characteristic
feature of junctions where the Josephson effect is
observed.}\label{Fig4}\end{center}
\end{figure}

As described earlier, in the experiments made in single grains, we
use the x-y table to change macroscopically the position of the
tip with respect to the sample. This enables us to clean the tip
on the gold surface and then to measure STM/S with a normal, clean
gold tip on a single MgB$_2$ grain. Sometimes however clear S-S
tunneling spectra appeared, due to a loosely connected single
grain which was attached by accident to the tip apex, as already
mentioned in \cite{Rubio01}, where we also reported that the
observed behavior was in agreement with the value of $\Delta$
found with a clean normal tip.

In Fig.4 we represent conductance curve obtained in one of those
experiments\cite{Rubio01}. A peak in the conductance due to the
Josephson current neatly appears at zero bias, and the so-called
subharmonic gap structure, characteristic of S-S junctions
(inset). It is indeed well known that multiple Andreev reflections
occurring at the junction provide a vehicle for a finite
conduction within the superconducting gap which becomes more
important when the resistance of the junction decreases (or its
transparency increases; in our experiments, we always measure
above $100k\Omega$, i.e. in the tunneling, and not in the point
contact regime, see \cite{Wolf}). This conduction mechanism is
enhanced when the bias voltage between both electrodes reaches
submultiples of two times the gap value, giving sharp features in
the conductance at $2\Delta/n$ (n=1,2,3,...), reminiscent of the
high quasiparticle peaks of the density of states\cite{Wolf}. In
Fig.4 we can clearly identify the dominant gap structure at
$2\Delta_{\pi}$ (with $\Delta_{\pi}=1.9mV$) and the peaks of the
subharmonic gap structure at $2\Delta_{\pi}/n$, with n=1,2,3,4
(higher order peaks fall below the resolution of the experiment).

On experiments in single crystals, we could also find S-S
tunneling spectra (Fig.5), corresponding to accidental situations
in which a loose MgB$_2$ piece was attached to the tip. We believe
that when the sample was broken before cooling the set-up, pieces
in good registry with the sample were probably cleaved out. Note
that the peak corresponding to a Josephson current is smaller than
the one in the previously discussed junction, possibly due to
differences in the Josephson coupling energy \cite{Wolf} (which
decreases with increasing tunneling resistance $R_N$; $R_N=
500k\Omega$ in Fig.5 and $R_N=150k\Omega$ in Fig.4; a Josephson
peak was also reported in SNS point contact junctions having a
resistance of about 30$\Omega$\cite{Zhang01})

\begin{figure}[btp]
 \begin{center}\leavevmode
 \includegraphics[width=0.8\linewidth]{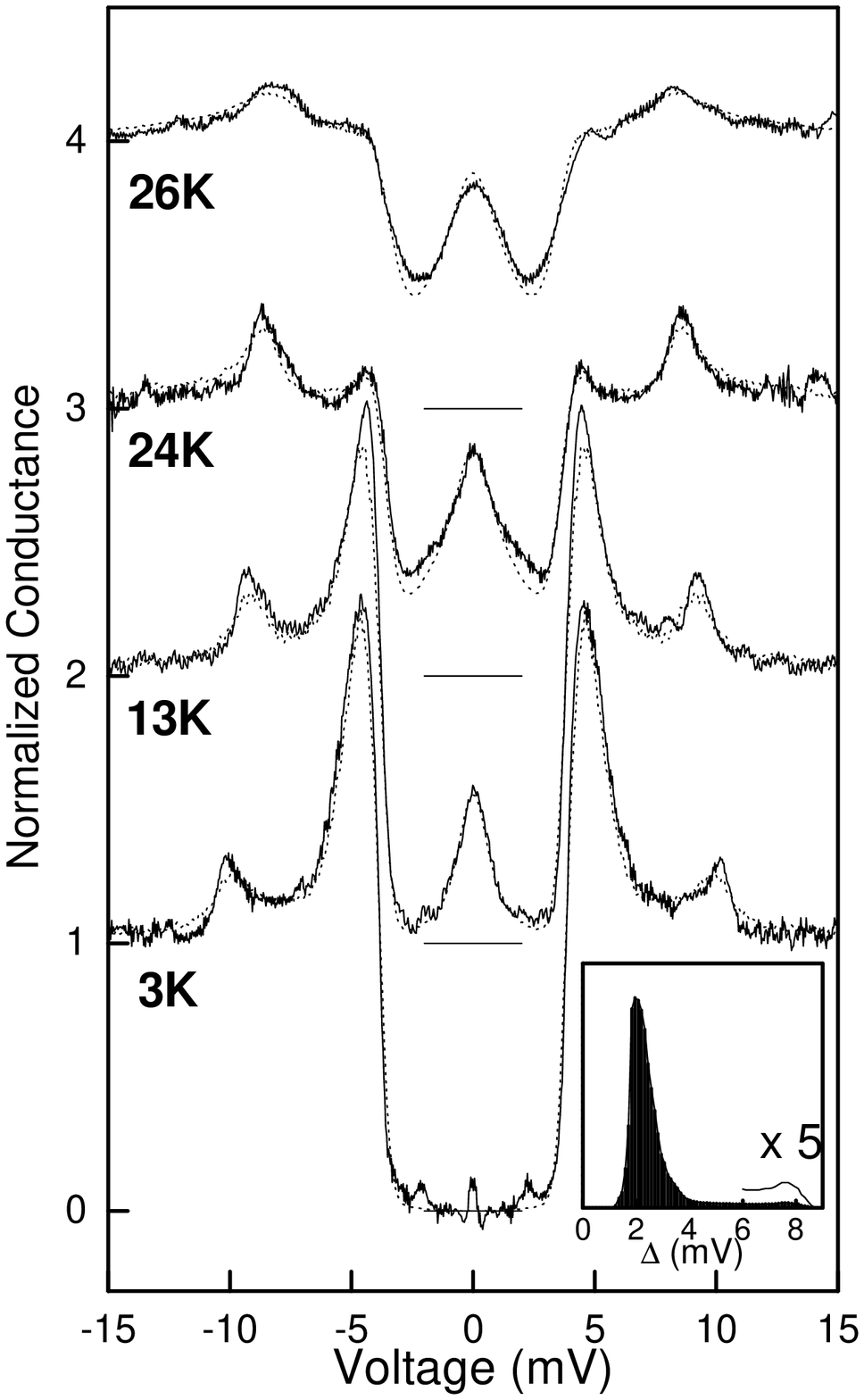}
\caption{The temperature dependence of the S-S spectra as measured
in a junction made of two parts of single crystalline MgB$_2$
(curves displaced by 1 in the y axis). Dashed lines are fits to
these curves. For the lowest temperature curve we use a
distribution of values of the superconducting gap as shown in the
inset at the bottom of the figure. For higher temperatures, this
distribution has to be changed slightly due to the temperature
variation of the superconducting gaps. The contribution from
$\sigma$ band electrons to the tunneling current is small and
leads to a broad peak, which has been zoomed out in the figure of
the inset. Straight lines at the bottom of each curve indicate
zero conductance for each temperature.}\label{Fig5}\end{center}
\end{figure}

\begin{figure}[btp]
 \begin{center}\leavevmode
 \includegraphics[width=0.8\linewidth]{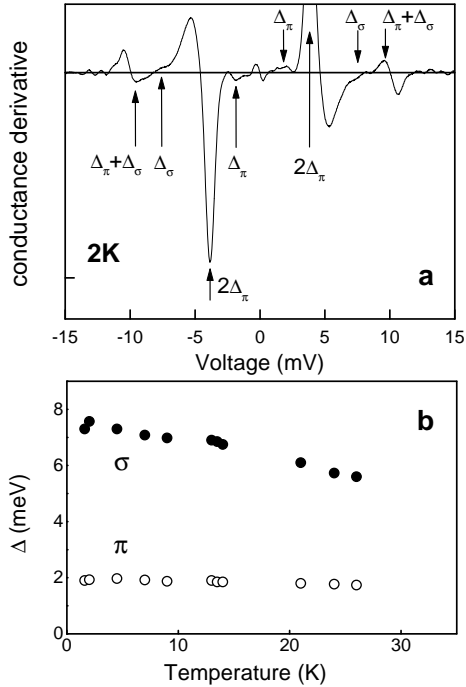}
\caption{The derivative of the conductance as a function of the
bias voltage in the S-S junction at 2K(a). This type of plots is
used to identify the voltage position of the main peaks obtained
in the conductance (Figs. 4 and 5). Note that the voltage range is
much larger than the one in inset of Fig.4, so that the peaks of
the subharmonic gap structure of $2\Delta_{\pi}$ corresponding to
$n>2$ (located at $2\Delta_{\pi}/n$, see right bottom inset of
Fig.4) appear as a very small feature at the center of this plot.
The arrows show the peaks which are best identified in this
voltage range. In (b) we show the temperature dependence of
$\Delta_{\sigma}$ and $\Delta_{\pi}$ as determined by using plots
as (a).}\label{Fig6}\end{center}
\end{figure}

The conductance of this junction as a function of temperature can
be fitted up to 26K to a distribution similar to the one shown in
the inset of Fig.5 for the 3K curve. Note that, in agreement with
Refs.\cite{Schmidt02,Zhang01}, we observe a very high peak at zero
bias when we increase the temperature. The combination of a large
amount of thermally excited quasiparticles and a superconducting
gap with a value much smaller than expected from most simple BCS
theory produces these anomalous curves.

In addition to the already discussed subharmonic gap structure and
the Josephson effect, two additional clear peaks appear at higher
bias voltages in Figs.4 and 5. Using the current procedure of
taking the derivative of the conductance as a function of the bias
voltage $d(\frac{dI}{dV})/dV$ \cite{Wolf}, which enhances the
features due to the dominating gaps in the distribution, we can
best identify the characteristic energy corresponding to the
physical process which gives these peaks, as a maximum (minimum)
in $d(\frac{dI}{dV})/dV$ for positive (negative) voltages. The
features at 3.8 and 9.5 mV are clearly due to conventional
tunneling processes and are also obtained from the fit to the
experimental curves shown in Figs.4 and 5. The feature in Fig.6a
at 3.8mV (corresponding to the highest peak in Figs.4 and 5), is
located at two times the dominant gap in the distribution
corresponding to the $\pi$ band gaps, $2\Delta_{\pi}$. The
contribution to the tunnelling current from the $\sigma$ band
electrons is too small in these junctions to give a feature at
$2\Delta_{\sigma}$ as shown by the fit in Fig.5 and its associated
gap distribution. The clear feature observed at 9.5mV is located
at the sum of the dominant gaps of each set of bands,
$\Delta_{\pi}+\Delta_{\sigma}$. The physical origin of this effect
is that, when the bias voltage between both electrodes coincides
with the sum of the peaks of both distributions
$\Delta_{\pi}+\Delta_{\sigma}$, the high peak in the density of
states corresponding to the $\pi$ band gaps enhances the small
bump corresponding to the $\sigma$ band gaps and gives a clearly
measurable peak in the conductance.

From Fig.6a we can also obtain the location of the peaks related
to the subharmonic gap structure associated to the $\Delta_{\pi}$
gap, already discussed in connection with Fig.4 (in Fig.6a, for
clarity, we only highlight $2\Delta_{\pi}/n$ with $n=2$).
Interestingly, we can also observe another feature at
$\Delta_{\sigma}$, possibly related to multiple Andreev
reflections which enhance the conductance at the voltage
corresponding to the $n=2$ subharmonic gap structure associated to
the $\sigma$ band gaps. This point should be analyzed in more
detail by extending presently available theoretical models to
multiband materials. To the best of our knowledge, no other
previous measurement of this characteristic feature of S-S tunnel
junctions has been made in multigap superconductors.

By using the same plot for higher temperature curves, we can also
easily obtain the temperature dependence of the dominant
superconducting gap features at the $\pi$ and $\sigma$ bands. The
result is represented in Fig.6b. Note that, while $\Delta_{\pi}$
stays fairly constant, $\Delta_{\sigma}$ changes more rapidly as a
function of temperature.

\section{Conclusion}

We have presented new results on Scanning Tunneling Spectroscopy
in single crystals and grains of MgB$_2$. The tunneling spectra
clearly reflect the multiband character of superconductivity in
this material, and are well fitted by two band BCS expressions. We
mostly find features associated to the $\pi$ band gap, which
clearly dominates the contribution to the tunneling spectra,
although features due to the $\sigma$ band gaps were also
detected. Both bands have a non negligible distribution of values
of the superconducting gap, in agreement with the recent
theoretical calculations of \cite{Choi02}.

In some locations of the surface we can observe interesting
effects with which we can extend our present understanding of this
material. Interband scattering effects lead to the observation of
a single gap structure. S-S junctions show clearly the Josephson
effect and the appearance of peaks at sub-multiples of the
superconducting gap characteristic of this type of junctions
(subharmonic gap structure).

Clearly, many questions remain open which need to be addressed in
future experiments. The observation of atomic resolution images of
the surface is necessary to understand the dominant role of the
$\pi$ band gap in our measurements. It is indeed striking that
other important experiments, as e.g. NMR or Raman scattering
\cite{Quilty02,Kotegawa01}, preferentially observe features
associated to the $\sigma$ band gap. In NMR \cite{Kotegawa01},
even no trace of the $\pi$ band gap has been measured. Although
these last experiments were made at a relatively high magnetic
field (1T), which could have influenced in some way the $\pi$ band
gap, the reasons for the observation of one or the other gap
structure in different experiments needs to be addressed.
Furthermore, the search for phonon features in the electronic
tunneling density of states in order to obtain more information
about the pairing interaction is an open issue. It should also be
interesting to study in more detail S-S junctions of this material
in a situation with a larger contribution from the $\sigma$ band
electrons, as for instance the one presented in fig.2b, to the
tunneling current to search for new features in the subharmonic
gap structure and in the Josephson effect. The resulting curves
should be highly non linear, opening the way to new interesting
applications.

Theoretical calculations pointing out possible new phenomena to be
observed in N-S and S-S tunneling experiments associated to the
multigap nature of superconductivity in this compound would be
very stimulating. In any case, MgB$_2$ is clearly a very
intriguing superconductor with a remarkably simple Fermi surface
and high critical temperature. This has made, and will continue to
make, feedback between theory and experiment especially fruitful.
We can therefore expect new and surprising physics to emerge from
this interaction in the following years.

\begin{ack}
We specially acknowledge F. Guinea and  E. Bascones for very
fruitful discussions. We also acknowledge J. Ramirez, N. Agrait
and J.L. Martinez. Support from the ESF programme Vortex Matter in
Superconductors at Extreme Scales and Conditions (VORTEX), as well
as from MCyT (MAT2001-1281-c02-01, Spain), the Comunidad
Aut\'onoma de Madrid (Spain) and the European Social Fund is
acknowledged. The single crystals of MgB$_2$ were prepared in
support by the New Energy and Industrial Technology Development
Organization (NEDO) as Collaborative Research and Development of
Fundamental Technologies for Superconductivity Applications.

\end{ack}


\end{document}